\documentclass[useAMS,usenatbib]{mn2e}

\usepackage{deluxetable}
\usepackage{graphicx}
\usepackage{times}

\title[Oxygen isotopic ratios in first dredge-up red giant stars and nuclear reaction rate uncertainties revisited]
      {Oxygen isotopic ratios in first dredge-up red giant stars and nuclear reaction rate uncertainties revisited}
\author[J. A. Stoesz and F. Herwig]{Jeffrey A. Stoesz$^{1}$$^{2}$ and Falk Herwig$^{1}$\thanks{E-mail: fherwig@uvastro.phys.uvic.ca}\\
$^{1}$Physics and Astronomy Department, University of Victoria, Canada\\
$^{2}$Herzberg Institute for Astrophysics, National Reseach Council, Victoria, Canada}

\begin{document}

\date{Submitted Sept 2002}

\pagerange{\pageref{firstpage}--\pageref{lastpage}} \pubyear{2002}

\maketitle

\label{firstpage}

\begin{abstract}
We describe a general yet simple method to analyse the propagation of nuclear
reaction rate uncertainties in a stellar nucleosynthesis and mixing
context. The method combines post-processing nucleosynthesis and
mixing calculations with a Monte Carlo scheme. With this approach we
reanalyze the dependence of theoretical oxygen
isotopic ratio predictions in first dredge-up red giant branch stars
in a systematic way. Such
predictions are important to the interpretation of
pre-solar Al$_{\rm 2}$O$_{\rm 3}$ grains from meteorites. 
The reaction rates with uncertainties were taken
from the NACRE compilation \citep{nacre}. We include seven reaction
rates in our systematic analysis of stellar models with initial masses
from 1 to 3$~{\rm M}_{\sun}$. We find that  
the uncertainty of the reaction rate for reaction ${\rm ^{18}O(p,\alpha)^{15}N}$
typically causes an error in the theoretical ${\rm ^{16}O/^{18}O}$ ratio of
$\simeq +20 /$$-5$ per cent.
The error of the  ${\rm ^{16}O/^{17}O}$ prediction is 
10--40 per cent depending on the stellar mass, and is
persistently dominated by the comparatively small uncertainty of the  ${\rm
^{16}O(p,\gamma)^{17}F}$ reaction. With the new estimates on reaction
rate uncertainties by the NACRE compilation, the p-capture reactions
${\rm ^{17}O(p,\alpha)^{14}N}$ and ${\rm ^{17}O(p,\gamma)^{18}F}$ have
virtually no impact on theoretical predictions for stellar mass $\leq 1.5~{\rm M}_{\sun}$.
However, the uncertainty in ${\rm ^{17}O(p,\alpha)^{14}N}$ has an effect comparable to or greater than that of ${\rm ^{16}O(p,\gamma)^{17}F}$ for masses $> 1.5~{\rm M}_{\sun}$, where core mixing and subsequent envelope mixing interact. 
In these cases where core mixing complicates post-dredge-up surface abundances, uncertainty in other reactions have a secondary but noticeable effect on surface abundances.
\end{abstract}


\section{Introduction}
Observations of abundances and abundance ratios in stars yield
powerful constraints on models of internal processes in stars.
In particular, the predicted abundance evolution of stars of a given
mass and metallicity depends on both mixing and nuclear burning
processes. 
While it is customary to provide observational results with
some estimate or analysis of the associated errors, this is seldomly
done for theoretical abundance predictions. This is partly due to the
fact that quantitative uncertainties on the input physics is not
available and that theoretical model uncertainties (in
particular for poorly understood mixing processes) are hard to
make. Finally, a systematic error propagation in stellar models can be
computationally demanding and the effort has often not been justified
in the past. However, new spectroscopic observations with unprecedented
precision will be available in the near future, and one has the
impression that stellar physics in general is entering a new
high-precision era. In this context it seems necessary to reconsider the
question of abundance prediction uncertainties in a quantitative way.

As an initial step in this direction, we have chosen
specifically the dependence of oxygen isotopic 
ratio predictions in giant stars on nuclear reaction rate
uncertainties. The oxygen isotopic ratios are of particular interest
because they are affected not only by nucleosynthesis but also by
mixing processes which are not yet very well understood. The latter is
evident from the spectroscopic properties of giant stars which are
not entirely consistent with standard stellar evolution models
\citep{harris85,harris:87,boothroyd:99}.  
Therefore it is important to quantitatively know
the uncertainties arising from nucleosynthesis in order to help 
constrain mixing processes using observations. In addition, the oxygen
isotope predictions are important for the interpretation of pre-solar
corundum (Al$_{\rm 2}$O$_{\rm 3}$) from meteorites for which oxygen
isotopic ratios can be measured with high precision
\citep[e.g.][]{huss,nittler97}.

\begin{figure}
\resizebox{\hsize}{!}
{\includegraphics[bbllx=30pt,bblly=400pt,bburx=600pt,bbury=700pt]{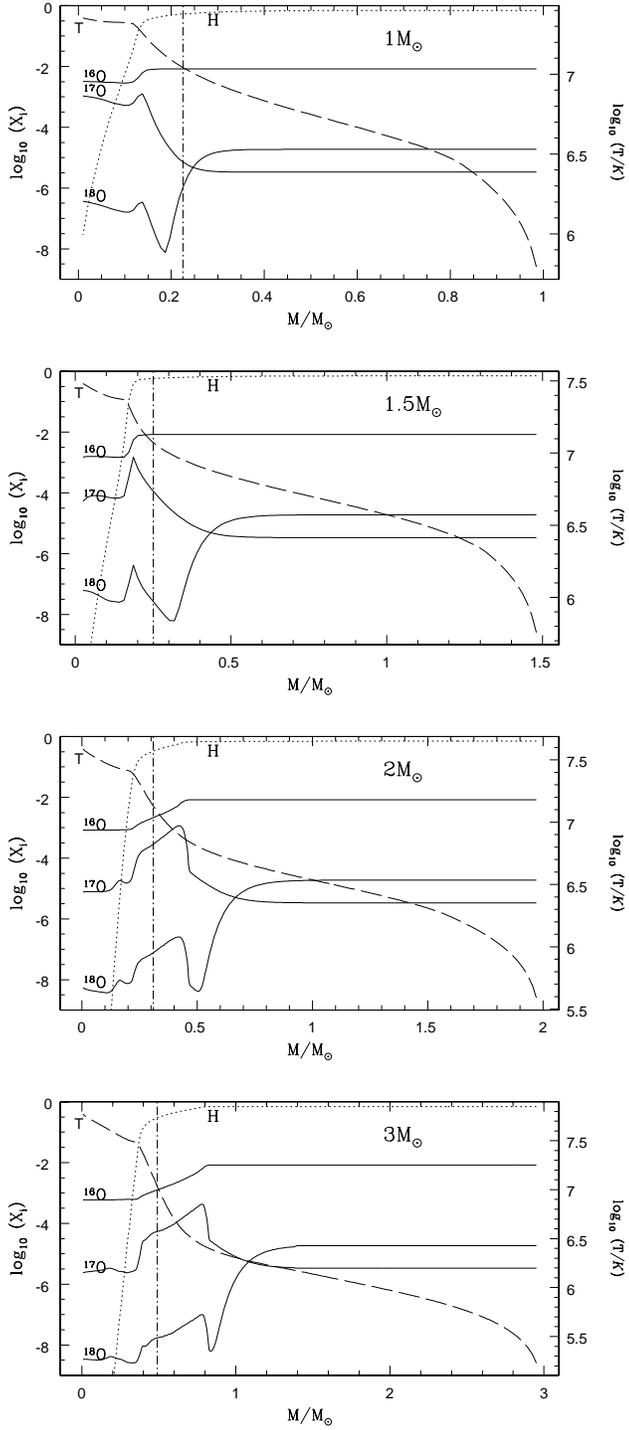}}
\caption{Mass fraction profiles of H (dotted), $^{16}$O, $^{17}$O, and $^{18}$O (solid), as well as temperature (long dash) for the $1~{\rm M}_{\sun}$ (top left), $1.5~{\rm M}_{\sun}$ (top right), $2~{\rm M}_{\sun}$(bottom left), and $3~{\rm M}_{\sun}$ (bottom right) cases with Z=0.02 near the end of the core H-burning. Also shown is the deepest penetration of the first dredge-up (dash-dot).}
\label{fig:abund}
\end{figure}

During the main sequence (MS) evolution of low- and
intermediate-mass stars, partial H-burning in the envelope produces $^{17}$O,
$^{13}$C, and $^{14}$N while $^{12}$C, $^{15}$N, and $^{18}$O are
destroyed (Fig.\,\ref{fig:abund}).
The abundance of $^{16}$O in the envelope is essentially unchanged for stellar mass $\leq 1.5~{\rm M}_{\sun}$.
For more massive stars, core convection on the early MS reaches out to $\sim 25$ per cent of the stellar mass, leaving behind core processed material as it retreats, including destroyed $^{16}$O \citep[e.g.][]{schaller:92}.
Evidence of the retreating core convection is seen in the 2 and 3$~{\rm M}_{\sun}$ cases in Fig.\,\ref{fig:abund} at the bottom of the envelope where the H and $^{16}$O are depleted.

At the end of the MS phase hydrogen is exhausted in the core, causing
the core to contract and the envelope to expand and cool. 
Envelope convection descends into the star as the envelope temperature drops.
Regions previously affected by cool H-burning are homogenised, leading
to a change of surface abundances (1$^{st}$ dredge-up, 1dup hereafter).
As a result of this mixing, theory predicts the $^{16}$O/$^{17}$O ratio decrease from
initially $\sim 2600$ to a few hundred, depending on stellar mass
while the $^{16}$O/$^{18}$O ratio increases only marginally by
$\leq 20$ per cent from the initial value of about 500, weakly dependent on mass.
For many stars, this is in fair agreement with spectroscopic
observations. For example, \citet{harris84} find for K-giants ${\rm
^{16}O/^{17}O}$ ratios in the range $300 \dots 1000$ and ${\rm
^{16}O/^{18}O} \sim 425 \dots 600$. Similarly, some corundum oxygen
isotopes are in rough agreement with standard 1dup model predictions
(e.g.\ SEAL203 in \citet{choi:98} has  ${\rm^{16}O/^{17}O}=355$ and a
solar   ${\rm ^{16}O/^{18}O}$ ratio) while many others require either
non-solar initial ${\rm ^{16}O/^{18}O}$ ratios \citep{huss,boothroyd94,timmes95} or non-standard mixing processes.

The dependence of oxygen isotopic ratios on the ${\rm^{17}O}$ proton
capture nuclear reactions has been studied before by \citet{eleid} and
\citet{boothroyd94}.
They considered the effect of a new determination of these reactions 
by \citet{landre:89} and found that the uncertainty of the  ${\rm
^{16}O/^{17}O}$ predictions is dominated by the uncertainty of these
rates for stars with initial masses larger than $\sim 2.5{\rm M_{\sun}}$.

In this paper we include seven reactions in a systematic analysis of the
influence of rate uncertainties  on the predicted oxygen isotopic
ratios in red giant branch (RGB) stars. In addition to addressing this scientific question, we also want to demonstrate and test the
method of evaluating model error propagation described here.
Sect. 2 describes the method and computations.
In Sect. 3 the uncertainty in the modeled oxygen isotopic ratios of
RGB stars is analysed for significant reaction rate uncertainties and
comparisons are made to observations and other models. 
A summary is in the final section.

\section{Method and Models}

The nucleosynthesis and mixing considered here do not alter
the structural evolution. Therefore we can efficiently
post-process a time sequence of full stellar models to analyse the
behaviour of nucleosynthesis and mixing in the envelope under
variations of the input nuclear reaction rates.   
A sequence is post-processed many times, each time with a different
set of values for the reaction rates (within published uncertainty) to obtain a Monte Carlo (MC) estimate of error
propagation into the surface abundance predictions.
In the following subsections we describe the stages of this
method in more detail. 

\subsection{Evolution code}
A time sequence of stellar models is generated by the {\it EVOL}
code \citep[see][and references therein]{falk,bloecker95}. 
At each time step the four partial differential equations of stellar
structure are solved on a 1D Lagrangian grid, with the {\it OPAL}
opacity tables from \citet{opal96}. 
A standard nuclear reaction network, including the PP-chains and the CNO tri-cycle, is incorporated. 

The treatment of mixing, in particular the use of convective 
overshooting, is not essential to this study because we are interested
here in the differential impact of the uncertainties of nuclear reaction
rates. However, there is some agreement that
the hydrodynamical properties of convection inevitably result in some
turbulent mixing into the neighboring stable layers, defined by the
convective boundary where the acceleration of streams disappears.
Models of convection in the hydrodynamical framework predict that the
turbulent velocity field decays roughly exponentially
\citep[e.g.][]{xiong:85,freytag:96,asida:00}. However, the extent of
overshooting is not the same at all convective boundaries. This is
evident from hydrodynamical simulations, for example by
\citet{freytag:96} who find $f_{ov}=0.25$ for the shallow surface
convection of A-stars and $f_{ov}=1.0$ for white
dwarfs. \citet{herwig:97} found that $f_{ov}=0.02$ reproduces the
results of \citet{schaller:92} who fitted the observed width of the
main sequence with an instantaneous treatment of overshooting of
$0.2H_{p}$. 
This order of magnitude for core convection
overshooting was confirmed by 2D hydrodynamical models by
\citet{deupree:00}. For a situation with an extended, deep envelope,
like
during the first dredge-up, an estimate of overshooting efficiency is
more
difficult.  \citet{alongi:91} found that the position
of the RGB luminosity bump can be aligned with observation with an
envelope overshooting three times the value required for the core
overshooting \citep[see, however,][]{cassisi:97}. \citet{bloecker:98b} 
found that overshooting at the 
bottom of the solar convection zone in excess of $f_{ov}=0.07$
would cause too much lithium destruction \citep[see
also][]{schlattl:99}. In view of the ambiguous evidence we decided to use  
the same value as for the main sequence core convection at the bottom of the
descending envelope convection during the first dredge-up evolution.

For the core convection we use no overshooting in the 1$~{\rm M}_{\sun}$ cases,
$f_{ov}=0.008$ in the 1.5$~{\rm M}_{\sun}$ sequence and  $f_{ov}=0.016$
for the 2$~{\rm M}_{\sun}$ and 3$~{\rm M}_{\sun}$ cases.
The mixing efficiency in the envelope is chosen to be $f_{env}=
0.016$ for all cases. 
The Reimer's mass loss formulae \citep{reimers:75} with $\eta=0.5$ for
$M<1.7~{\rm M}_{\sun}$ and $\eta=1.0$ for $M>1.7~{\rm M}_{\sun}$ is employed,
starting at the bottom of the RGB.

\subsection{Post-processing code and the Monte Carlo Scheme}
\label{sec:pp-mc}
Our post-processing code takes the stellar structure from the full
stellar evolution model sequences generated by {\it EVOL} as input and
recomputes abundance changes due to nucleosynthesis and mixing. 
The code is fully implicit, iterative, and couples mixing with
nucleosynthesis using adaptive step size and full error control. 

All reactions in the post-processing code use NACRE reaction rates
\citep{nacre}, which include estimates on upper and lower limits. We
have used the electronic web tool \emph{Netgen} \citep{jorissen} to retrieve the tabulated rates.
The full network of nuclear reactions is the same as in the
full stellar evolution code, but  rates  for a
predefined set of seven reactions were systematically altered. 

If one is interested in knowing the abundance uncertainty of a
particular species, then inspection of its production and destruction reactions may reveal to which reaction uncertainty the abundance is most sensitive. 
For example, $^{18}$O is mainly destroyed rather than produced in the envelope.
Hence, the surface abundance of $^{18}$O after 1dup ought to be most sensitive to the one of the two major reactions which consume it.
Reaction ${\rm ^{18}O(p,\alpha)^{15}N}$ has a larger nominal rate than
${\rm ^{18}O(p,\gamma)^{19}F}$ and accordingly this reaction must be
most important to $^{18}$O. While such a qualitative consideration is
straightforward it is harder to estimate the
quantitative effect of the uncertainty of this reaction because the
combined effects of ${\rm ^{18}O(p,\alpha)}$ and the other reactions
when the temperature profile evolves before equilibrium is reached and
core convection on the early main sequence may affect the envelope abundances. 

Manipulating the rates of one reaction that is part of a
larger network will not necessarily reveal the whole picture, especially when
equilibrium abundances are not achieved. This is true in particular
for a general network situation in which reaction flows may be
redirected due to uncertainties of reactions that are involved in
branchings. In order to have a tool available which can readily
be applied to any nucleosynthesis and mixing scenario we
decided to take a general and flexible approach by combining
post-process calculations with an MC scheme, as described
below. With this approach we could, for example, also study more complex
situations like the dependence of the $s$-process in asymptotic giant branch (AGB) stars on
nuclear reaction rate uncertainties.  

For each MC iteration the reaction rate for reaction $x$ is given by
\begin{equation}
{\rm R_i^x(T)=\xi_i^x(R_u^x(T)-R_n^x(T))+R_n^x(T),~~~1\geq\xi_i^x\geq 0}
\end{equation}
\begin{equation}
{\rm R_i^x(T)=\xi_i^x(R_n^x(T)-R_l^x(T))+R_n^x(T),~~~0>\xi_i^x\geq -1}
\end{equation}
where ${\rm 1\geq\xi_i^x\geq -1}$, and ${\rm R_u^x,~R_n^x,~R_l^x}$ are the upper, nominal, and lower rates respectively.
We choose a Gaussian distribution to the random variable $\xi_i$ which peaks at zero and has an e-folding distance at $\xi_i=.55$ so that the extreme values are populated at 1.8$\sigma$. 
Note that while the distribution on $\xi_i$ is symmetric, the distribution of ${\rm R(T)}$ is not.

\subsection{Details of the computed models} \label{sec:details}

With the stellar evolution code we generated seven sequences of stellar
structure models for a range of initial mass and metallicity.
 Each sequence was followed from 
the beginning of the MS to just after the end of the 1dup. 
The stellar model sequences together with the mass coordinate of the
deepest penetration of the 1dup (${\rm M_{dup}}$) and the maximum
temperature achieved at that mass coordinate are
listed in Table\,\ref{tbl:cases}. 
The maximum envelope temperature was achieved
at about $\frac{4}{5}$ of the MS lifetime for all cases. 
The envelopes of the 2--3$~{\rm M}_{\sun}$ cases receive matter from the core due to core convection on the early MS.
For these cases, Table \ref{tbl:cases} also shows the maximum temperature achieved at the centre of the star ($T_c$) from the beginning of the MS to when the core convection no longer overlaps with the envelope.
These temperatures are the maximum that part of the surface material, after 1dup, was exposed to.

Out of the full stellar evolution model sequences about 32 to 36 models for 1--1.5$~{\rm M}_{\sun}$ cases and 82 to 119 models for 2--3$~{\rm M}_{\sun}$ were chosen for post-processing.
The bulk of extra models used in the 2--3$~{\rm M}_{\sun}$ cases were put on the first half of the MS, when core convection was retreating, leaving behind material processed in the core.
At least 4 models were put near the time
of deepest 1dup to ensure that  a precise mixing depth was achieved. 
We post-processed the whole star for 2--3$~{\rm M}_{\sun}$ but only the envelope layers affected by the 1dup for 1--1.5$~{\rm M}_{\sun}$.
In this study we ignore any extra mixing processes that might occur later
on the RGB, like cool bottom processing \citep[][]{wasserburg95}. 
The choice of stellar cases spans 1$~{\rm M}_{\sun}$ to 3$~{\rm M}_{\sun}$ with
solar metallicity and additional cases for 1$~{\rm M}_{\sun}$ and
2$~{\rm M}_{\sun}$ where the metallicity is lower. For the lower
metallicity cases we have scaled all three oxygen isotopes with Z,
{\it not} taking into account the anti-correlation of the ${\rm
^{16}O/^{18}O}$ with Z reported by \citet{timmes95}.
To facilitate comparison, the initial oxygen isotope ratios were chosen to be ${\rm ^{16}O/^{17}O_i=2465}$ and ${\rm ^{16}O/^{18}O_i=442}$, the same as in \citet{eleid}.

Fig. \ref{fig:abund} shows example abundance profiles produced by
the post-processing code using the nominal reaction rates, together with
the 1dup mark. 
One can see that $^{17}$O is particularly sensitive to the maximum
depth of the 1dup for the 1$~{\rm M}_{\sun}$ and 1.5$~{\rm M}_{\sun}$ cases because
of the sharp rise in $^{17}$O at the bottom of the 1dup \citep{landre90}. 
The reduced $^{16}$O at the bottom of the envelope, from where core convection retreated, enhances the sensitivity to the maximum
depth of the 1dup for the 2$~{\rm M}_{\sun}$ and 3$~{\rm M}_{\sun}$ cases.

\begin{figure}
\resizebox{\hsize}{!}
{\includegraphics[bbllx=40pt,bblly=50pt,bburx=520pt,bbury=440pt]{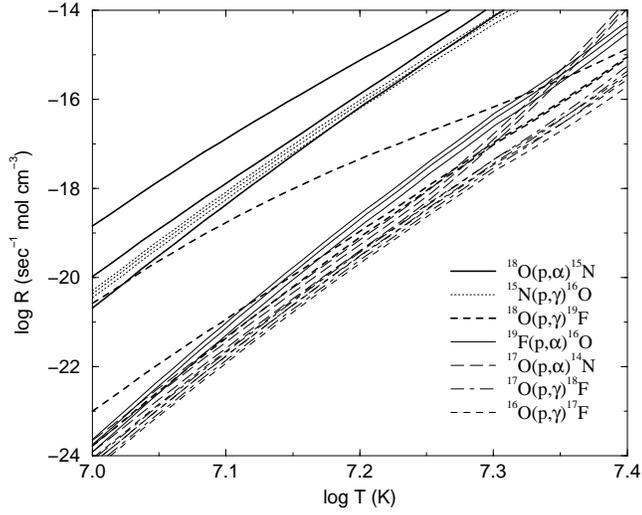}}
\caption{The nominal, upper and lower limits of the rates for the seven reactions. The order of reations in the legend is the same as Table \ref{tbl:reac}.}
\label{fig:rat}
\end{figure}

The seven CNO cycle reactions which are included in the analysis
are listed in Table\,\ref{tbl:reac} in decreasing order of ${\rm
R_n(\log{\rm T}=7)}$.  
The reaction ${\rm ^{16}O(p,\gamma)^{17}F}$ is followed by a rapid $\beta^+$ decay and
draws from a large reservoir of $^{16}$O to produce $^{17}$O.  
Reaction ${\rm ^{17}O(p,\gamma)^{18}F}$ consumes $^{17}$O to produce $^{18}$O.
Branching off from this series are reactions ${\rm ^{18}O(p,\alpha)^{15}N}$ and ${\rm ^{18}O(p,\gamma)^{19}F}$, which consume $^{18}$O, and ${\rm ^{17}O(p,\alpha)^{14}N}$, which consumes $^{17}$O.
For the 1$~{\rm M}_{\sun}$ cases, reaction ${\rm ^{16}O(p,\gamma)^{17}F}$ is slow enough and $^{16}$O is
 abundant enough that the $^{16}$O reservoir is
effectively not changing in the envelope. 
This is not true for 2--3$~{\rm M}_{\sun}$ (see Fig. \ref{fig:abund}),
and therefore reactions ${\rm ^{15}N(p,\gamma)^{16}O}$ 
and 
${\rm ^{19}F(p,\alpha)^{16}O}$
should receive attention in these
cases because they affect the production of $^{16}$O. 
Fig. \ref{fig:rat} shows the nominal, upper and lower limits on these reaction rates for the relevant temperature range. 

\section{Results}

The primary result from the MC simulations are quantified model uncertainties in the oxygen isotopic ratios due to reaction rate uncertainties. 
Secondly, we assess the individual contribution from each of the considered reaction uncertainties on isotopic ratio uncertainties.

An MC simulation was carried out for each of the cases in Table
\ref{tbl:cases} and the results are summarized in Table
\ref{tbl:uncer} and Fig. \ref{fig:tripleiso}. 
The isotopic ratio uncertainties are derived from the distribution of points in triple isotope plots.
The upper error bars for ${\rm ^{16}O/^{18}O}$ are defined to be the location of a horizontal line separates $2$ per cent of the points above the line, and $98$ per cent of the points are below.
The lower error bar is where $98$ per cent are above and $2$ per cent are below.
Error bars for ${\rm ^{16}O/^{17}O}$ are defined similarly. 
Therefore, the total number of MC points that lie outside of the error bars is $\leq 8$ per cent.
This scheme helps ensure that the errors bars are nearly invariant to the number of iterations.
There were 200 iterations for each 1--1.5$~{\rm M}_{\sun}$ case, and at least 400 iterations for each $> 1.5~{\rm M}_{\sun}$ case, which are already more time consuming due to a greater number of input models.
More iterations are needed for the $> 1.5~{\rm M}_{\sun}$ cases because there are 3 instead of 2 influential reaction rate uncertainties.

\subsection{The triple isotope plots from the MC simulation}\label{sec:priority}

\begin{figure}
\resizebox{\hsize}{!}
{\includegraphics[bbllx=20pt,bblly=150pt,bburx=570pt,bbury=700pt]{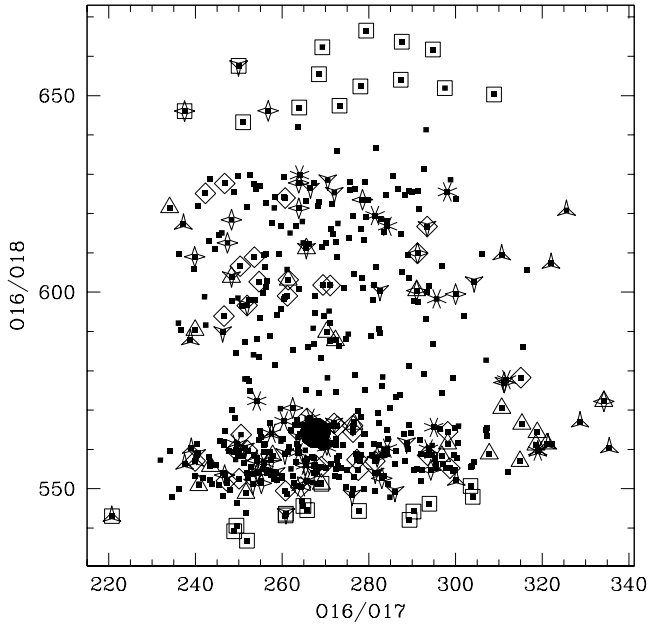}}
\caption{The surface oxygen isotope ratios for $3~{\rm M}_{\sun}$ and Z=0.02
from an MC simulation with all seven reactions and 500 iterations. All points are plotted in
small filled squares and over-plotted are points from iterations where
$|\xi_i^x|>0.7$ for each reaction; ${\rm ^{15}N(p,\gamma)^{16}O}$ (diamonds), ${\rm ^{16}O(p,\gamma)^{17}F}$ (open
triangles), ${\rm ^{17}O(p,\alpha)^{14}N}$ 
(starred triangles up), ${\rm ^{17}O(p,\gamma)^{18}F}$ (asterisk), ${\rm ^{18}O(p,\alpha)^{15}N}$ (open
squares), ${\rm ^{18}O(p,\gamma)^{19}F}$ (starred triangles down), ${\rm ^{19}F(p,\alpha)^{16}O}$ (starred diamonds). Note the
lack of open triangles and squares in the centre (error dominant reactions ${\rm ^{16}O(p,\gamma)^{17}F}$
and ${\rm ^{18}O(p,\alpha)^{15}N}$).}\label{fig:scatdemo} 
\end{figure}

\begin{figure}
\resizebox{\hsize}{!}
{\includegraphics[bbllx=20pt,bblly=150pt,bburx=570pt,bbury=700pt]{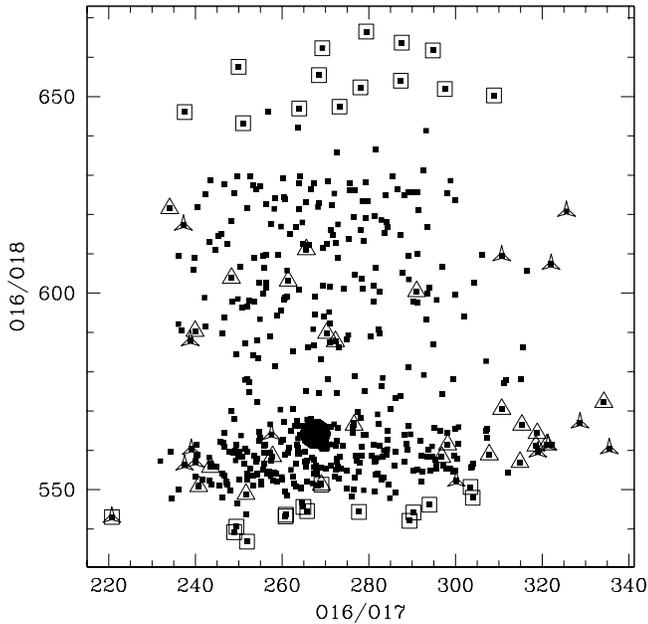}}
\caption{Same as Fig. \ref{fig:scatdemo} for the $3~{\rm M}_{\sun}$ case, except only symbols from iterations where $|\xi_i^{{\rm ^{16}O(p,\gamma)^{17}F}}|>0.7$ (open triangles), $|\xi_i^{{\rm ^{17}O(p,\alpha)^{14}N}}|>0.7$ 
(starred triangles up), and $|\xi_i^{{\rm ^{18}O(p,\alpha)^{15}N}}|>0.7$ (open squares) are over plotted. A filled circle marks the point where the nominal rates were used for all reactions.}
\label{fig:scat}
\end{figure}

Fig. \ref{fig:scatdemo} shows a typical result from an MC simulation.
Each point on the triple oxygen isotope plot represents the
isotopic ratios after the 1dup for one iteration.
The set of $\xi_i^x$ for each iteration allows one to locate points
with reaction rates far from the nominal value.
In Fig. \ref{fig:scatdemo} a symbol associated with  reaction $x$ is
over-plotted if $|\xi_i^x|>0.7$ (i.e. $>1.4\sigma$) is satisfied for that reaction.
Multiple symbols may be over-plotted for individual iterations.
In this way the oxygen isotopic ratios from iterations for which reaction
$x$ was near an upper or lower limit are flagged. The distribution
of the flagged iterations can reveal the relative importance of the
uncertainties of the reactions considered.
Symbols for reactions that do not dominate the isotopic
uncertainties can appear everywhere in the cloud of points, indicating that a rate far from the nominal value
for these reactions does not cause any preferred location in the plot (reaction ${\rm ^{15}N(p,\gamma)^{16}O}$ for example).
The lack of symbols from reaction ${\rm ^{18}O(p,\alpha)^{15}N}$ in a horizontal band through the central region of the plot indicates that its reaction rate uncertainty overwhelmingly dominates the others in affecting the ${\rm ^{16}O/^{18}O}$ error. 
Reaction ${\rm ^{18}O(p,\alpha)^{15}N}$ causes scatter on the ordinate because it destroys $^{18}$O.
Fig. \ref{fig:scat} shows this more clearly.

\begin{figure}
\resizebox{\hsize}{!}
{\includegraphics[bbllx=50pt,bblly=390pt,bburx=540pt,bbury=720pt]{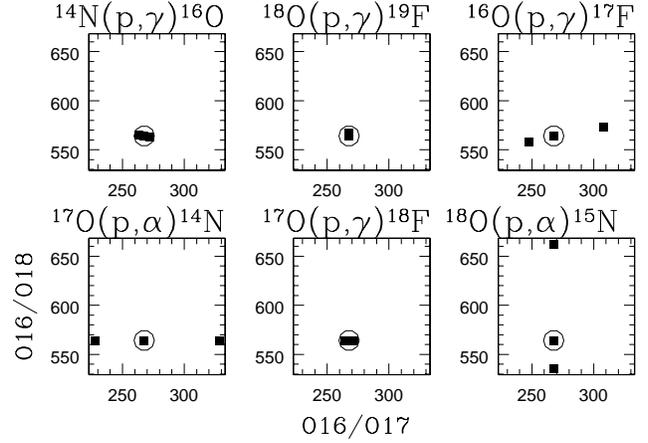}}
\caption{Oxygen isotope ratios for the 3$~{\rm M}_{\sun}$ Z=0.02 case when one reaction took the upper, lower, and nominal (circled) reaction rate while the other rates remained at nominal. Results from ${\rm ^{19}F(p,\alpha)^{16}O}$ (not shown) are the smallest.}
\label{fig:isodemo}
\end{figure}

\begin{figure}
\resizebox{\hsize}{!}
{\includegraphics[bbllx=50pt,bblly=390pt,bburx=540pt,bbury=720pt]{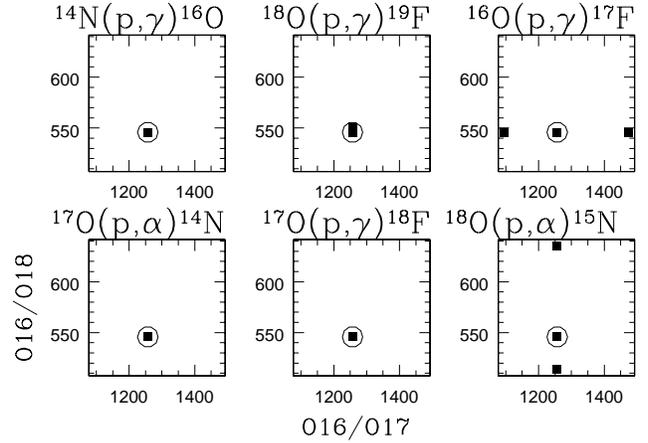}}
\caption{Same as Fig. \ref{fig:isodemo}, for the 1.5$~{\rm M}_{\sun}$ Z=0.02 case.}
\label{fig:isodemo1.5}
\end{figure}

\begin{figure}
\resizebox{\hsize}{!}
{\includegraphics[bbllx=20pt,bblly=150pt,bburx=570pt,bbury=700pt]{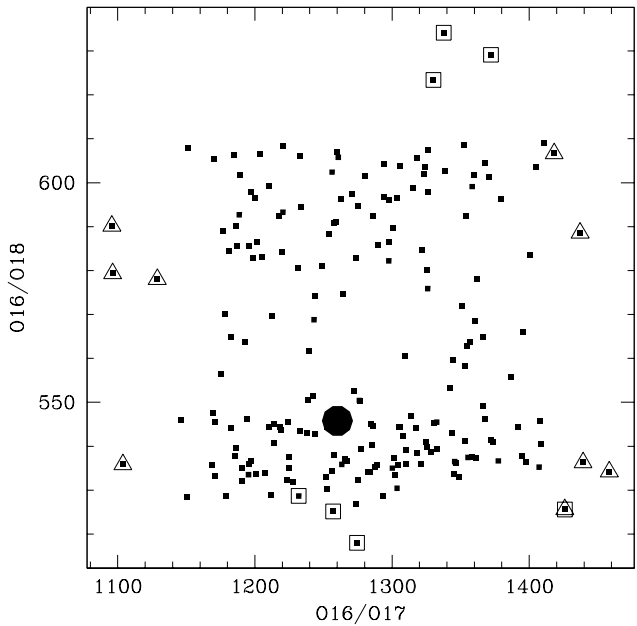}}
\caption{Same as Fig. \ref{fig:scatdemo} for the $1.5~{\rm M}_{\sun}$ case with 200 iterations and only symbols from iterations where $|\xi_i^{{\rm ^{16}O(p,\gamma)^{17}F}}|>0.7$ (open triangles) and $|\xi_i^{{\rm ^{18}O(p,\alpha)^{15}N}}|>0.7$ (open squares) are over plotted. A filled circle marks the point where the nominal rates were used for all reactions.}
\label{fig:scat15}
\end{figure}

For the 3$~{\rm M}_{\sun}$ case shown, reactions ${\rm ^{16}O(p,\gamma)^{17}F}$ and ${\rm ^{17}O(p,\alpha)^{14}N}$ are responsible for the scatter on the abscissa because they most strongly affect $^{17}$O.
This fact is not clear from Fig. \ref{fig:scat} because neither one clearly dominates.
With random $\xi_i$ there are many iterations where there is a cancelling effect, leading to points in the middle part of the cloud.
In order to demonstrate this we do 15 separate runs of the post-processing code. 
In the first run, all rates are held at nominal.
In the subsequent 14 runs, each of the 7 reaction rates takes its upper, then lower value while the other rates remain at the nominal value.
The result of the 15 runs is illustrated in Fig. \ref{fig:isodemo} for the $3~{\rm M}_{\sun}$ case.
Reaction ${\rm ^{17}O(p,\alpha)^{14}N}$ affects the abscissa and reaction ${\rm ^{16}O(p,\gamma)^{17}F}$ affects the ordinate secondly, but primarily affects the abscissa, as one might expect from inspection of the CNO network.
Fig. \ref{fig:isodemo1.5} shows a similar 15 combination run on the 1.5$~{\rm M}_{\sun}$ case.
For masses $\leq 1.5~{\rm M}_{\sun}$ the envelope does not receive matter processed in the core, and hence the surface abundances are a product of nuclear processing at cooler temperatures. 
As a result, uncertainty of the ${\rm ^{17}O(p,\alpha)^{14}N}$ reaction loses all of its effect on $^{16}$O/$^{17}$O uncertainty for stellar mass $\leq 1.5~{\rm M}_{\sun}$ cases.
Fig. \ref{fig:scat15} shows the absence of points in the centre for the dominant reactions.

The point distribution in the MC triple isotope plots of course depends on the choice of the e-folding distance for the random variable $\xi_i$.
The distributions typically have patterns due to the asymmetric nature of reaction rate
uncertainties. For example the bimodal distrubution in
Fig.\,\ref{fig:scat} (high point density in the lower half and low
density for larger  $^{16}$O/$^{18}$O) is a consequence
of the large upper limit for reaction ${\rm ^{18}O(p,\alpha)^{15}N}$.  

\subsection{Discussion of reaction uncertainties} \label{sec:discuss}

Reaction ${\rm ^{16}O(p,\gamma)^{17}F}$ has a relatively low  uncertainty (Fig.\,\ref{fig:rat}). For
example at $\log T=7.4$ the upper and lower limit according to the
NACRE compilation are $\pm 30$ per cent. This reaction
derives its effect from the large (virtually constant) reservoir of
$^{16}$O which leads to a strong production rate of $^{17}$O. 
Although the reaction rate of ${\rm ^{17}O(p,\alpha)^{14}N}$ is about
two times greater than that of ${\rm ^{16}O(p,\gamma)^{17}F}$, $^{17}$O
is produced because the abundance of $^{16}$O is about 3000 times greater than that of $^{17}$O. Hence, the uncertainty in reaction ${\rm
^{16}O(p,\gamma)^{17}F}$ is more important to the surface abundance of
$^{17}$O.\footnote{Production (or destruction) rate is
$r=N_aN_b\langle\sigma \upsilon\rangle$, where $N_a$ and $N_b$ are the
molar densities of the reactants and $\langle\sigma \upsilon\rangle=R$
is the reaction rate.} 
In the past, reaction rates for ${\rm ^{17}O(p,\alpha)^{14}N}$ and ${\rm ^{17}O(p,\gamma)^{18}F}$ were very uncertain \citep[see Sect. 3.2][]{eleid}.
With the modern, smaller NACRE uncertainty, uncertainty in reaction ${\rm ^{17}O(p,\alpha)^{14}N}$ is still relevant to the
$^{17}$O/$^{16}$O 1dup predictions for the 2--3$~{\rm M}_{\sun}$ cases.
Modern uncertainty in reaction ${\rm ^{17}O(p,\gamma)^{18}F}$, however, is not significant in the cases studied here.

For the 2--3$~{\rm M}_{\sun}$ cases, reactions ${\rm ^{17}O(p,\alpha)^{14}N}$ and ${\rm
^{16}O(p,\gamma)^{17}F}$ (starred triangles and triangles respectively in Fig. \ref{fig:isodemo}) have a similar influence on $^{16}$O/$^{17}$O.  
In the 2--3$~{\rm M}_{\sun}$ cases the combined effect of the uncertainty in these two reaction rates can cooperate to give some extreme $^{16}$O/$^{17}$O values as well as partially cancel to give points in the central part of the cloud. 

Reaction ${\rm ^{18}O(p,\alpha)^{15}N}$ strongly effects the isotopic abundances because its
uncertainty is very large at the envelope temperatures of these stellar
cases.  
Also note that reaction ${\rm ^{18}O(p,\gamma)^{19}F}$ has no noticeable effect on $^{18}$O.
Even though the error for this reaction is very large, reaction ${\rm ^{18}O(p,\alpha)^{15}N}$ is
faster and therefore more important to the destruction of $^{18}$O.

The effect of stellar mass and metallicity on the isotope ratio uncertainties is detectable mostly in $^{16}$O/$^{17}$O. 
The $^{16}$O/$^{18}$O uncertainty is typically $+20/-5$ per cent due to reaction ${\rm ^{18}O(p,\alpha)^{15}N}$.
For cases $>1.5~{\rm M}_{\sun}$, with higher temperatures (2$~{\rm M}_{\sun}$, Z=0.001 and 3$~{\rm M}_{\sun}$, Z=0.02 especially) greater $^{16}$O destruction and $^{17}$O production can be seen at the surface.
Hence, the destruction rate of $^{17}$O rivals its production from $^{16}$O, so the effect of uncertainty in reaction ${\rm ^{17}O(p,\alpha)^{14}N}$ is visible at the surface.

The interaction between the core and envelope convection for the $>1.5~{\rm M}_{\sun}$ cases means different temperature environments are seen by some of the dredged up material, than are seen by the $\leq 1.5~{\rm M}_{\sun}$ cases.
This adds to the complexity and has a very noticeable impact on the propogation of reaction rate uncertainty to surface abundances because it means uncertainties at a wider variety of temperatures become important.

\subsection{Comparison with observations and other models}

\begin{figure}
\resizebox{\hsize}{!}
{\includegraphics[bbllx=40pt,bblly=50pt,bburx=520pt,bbury=440pt]{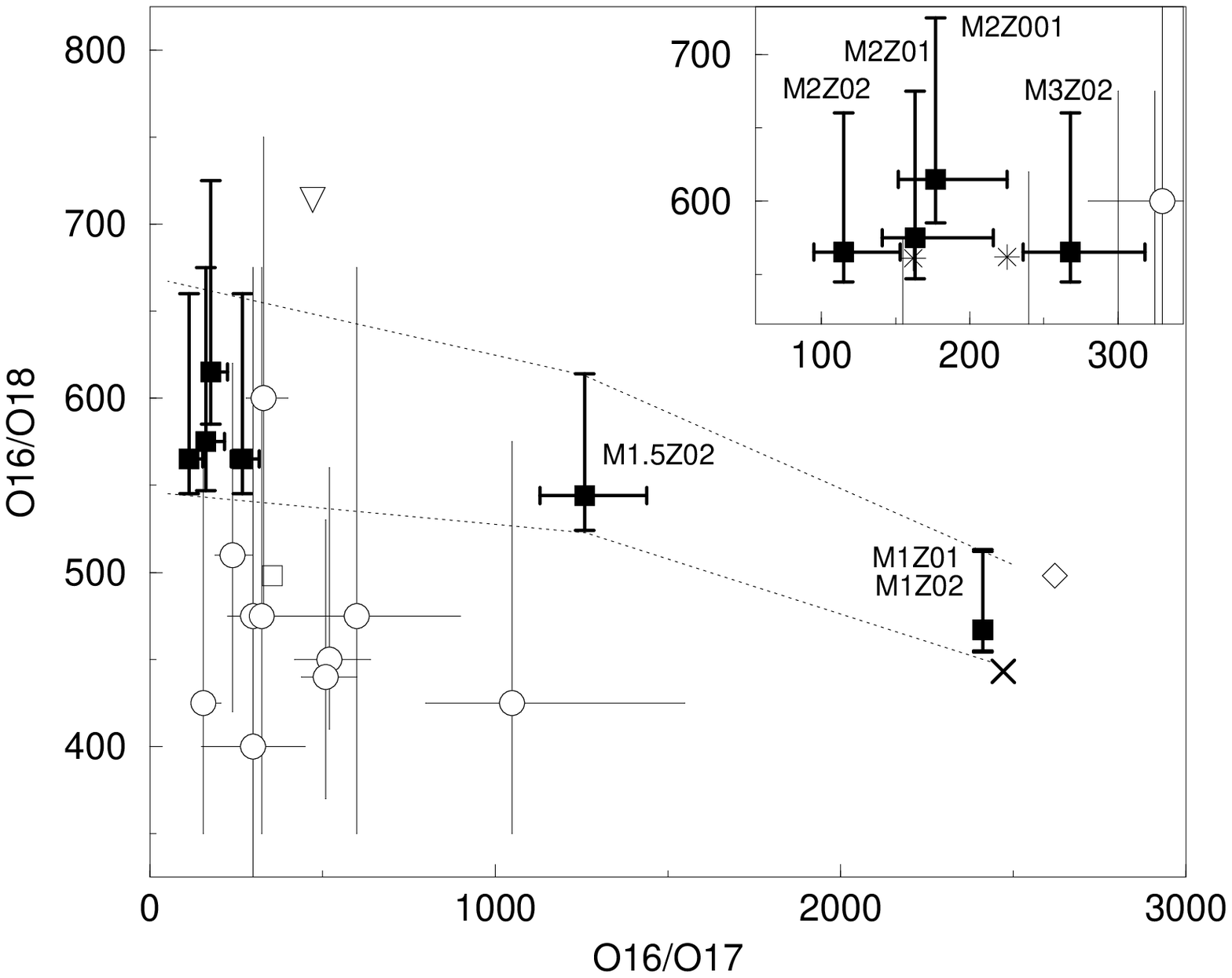}}
\caption{Modeled isotopic ratios with propagated errors
derived from MC simulations for the seven stellar cases (filled squares, labeled with their mass and metallicity). The x marks the initial oxygen ratios before 1dup for all cases. Spectroscopic data (circles) from
\citet{harris84, harris88}. Meteoric data from
SEAL203 (open square), SEAL235 (open triangle) \citep{choi:98}, and solar (diamond). Dotted lines mark a band where Z=0.02 stars are predicted to exist with our choice of initial abundances. Inset shows our 2--3$~{\rm M}_{\sun}$ model predictions in more detail, as well as predictions of \citet{eleid} for Z=0.02, 2$~{\rm M}_{\sun}$ (left asterisk) and 3$~{\rm M}_{\sun}$ (right asterisk). }
\label{fig:tripleiso}
\end{figure}

Our isotopic ratios, with the nominal rates, agree with results from \citet{boothroyd94} that predict a strong dependence of $^{16}$O/$^{17}$O on stellar mass. 
With increasing stellar mass from 1$~{\rm M}_{\sun}$ to 2$~{\rm M}_{\sun}$,
$^{16}$O/$^{17}$O decreases from 2500 to 100 for Z=0.02 in
Fig. \ref{fig:tripleiso} (i.e. $^{16}$O destruction in the envelope
increases for $>1.5~{\rm M}_{\sun}$ and the $^{17}$O production is increased
in all cases).    
Our predictions show increasing $^{16}$O/$^{17}$O with increasing mass for the $>2~{\rm M}_{\sun}$ cases, producing a minimum at about 2$~{\rm M}_{\sun}$, the same as in \citet{eleid} \citep[see also][]{boothroyd94}.
The inset in Fig. \ref{fig:tripleiso} shows model predictions by \citet{eleid} and this study (asterisks and filled boxes respectively) for 2$~{\rm M}_{\sun}$ and 3$~{\rm M}_{\sun}$.
The difference between our $^{16}$O/$^{17}$O predictions and those of \citet{eleid} is partially explained by our different choice for the important ${\rm ^{17}O(p,\alpha)^{14}N}$ reaction rate.
The lower limit on this rate, used in this study, is similar to the rate used to predict the asterisks.
Uncertainty in the treatment of mixing processes, which is not propagated with the MC simulations, likely makes the largest contribution to the difference.

The dependence of oxygen abundance ratios on stellar metallicity is tied to dependence of structure (temperature) to metallicity.
The predicted ratios for the two 1$~{\rm M}_{\sun}$ cases are indistinguishable given the calculated uncertainty.
The predictions for three 2$~{\rm M}_{\sun}$ cases are distingushable.
Lower metallicity produces higher temperatures and greater overall $^{18}$O destruction, causing higher $^{16}$O/$^{18}$O for those cases.

Spectroscopic observations of the isotope ratios in RGB stars from \citet{harris84, harris88} are also shown in Fig. \ref{fig:tripleiso} and Table \ref{tbl:obs}.
These stars are either ascending the RGB for the first time or are undergoing core He-burning on the blue loops.
Hence, the surface abundances of these stars have been changed by the 1dup.
The mass-$^{16}$O/$^{17}$O relationship is roughly obeyed by these observatations \citep[see][for a discussion]{eleid}.

The prediction band in Fig. \ref{fig:tripleiso} was from an initial $^{16}$O/$^{18}$O which is 10 per cent smaller than the solar value, giving predictions that overlap with all but one of the spectroscopic observations.
However, the number of post-1dup observations below solar ($^{16}$O/$^{18}$O=498) and below the band predicted from initial $^{16}$O/$^{18}$O=442 is worth noticing.
When there is a slight shift in the initial oxygen isotopic ratios, there is a proportional shift in the modeled post-1dup ratios \citep[as shown by][]{boothroyd94}. 
Hence, if the model prediction of about a $15$ per cent increase in $^{16}$O/$^{18}$O after 1dup is correct, then some of the stars with spectroscopic data in Fig. \ref{fig:tripleiso} (circles) must have had initial $^{16}$O/$^{18}$O values $\sim 30$ per cent less than solar (super-solar metallicity). 
In addition, the post-1dup spectroscopic data shown in Fig. \ref{fig:tripleiso} roughly correlate with [Fe/H] measurments compiled by \citet{taylor} (Table \ref{tbl:obs}), but indicate a pre-solar metallicity. 
There is currently no reason to believe that $^{18}$O is not depleted in the envelope during the MS.
A possible explaination is additional mixing above the core that further depletes $^{16}$O in the envelope, leading to an decreased $^{16}$O/$^{18}$O after 1dup.
Testing this hypothesis will be appropriate when more precise spectroscopic observations become available.


Pre-solar meteoritic inclusions have been linked
to giant stars due to their specific
isotopic abundance signatures \citep{huss,zinner}.
The rather large spread of their oxygen isotopic ratios implies additional mechanisms at work (like intital isotope ratio variations
or extra mixing processes). The variations of isotope ratios in grains
are much larger than the model uncertainties due to nuclear reaction
rates.
However, better nuclear data would be required to improve the
identification
of these extra mechansims.



\section{Conclusions}

In this paper we quantify the uncertainty in oxygen isotopic ratios due to uncertain reaction rates.
These results may help to motivate and prioritize new laboratory measuremtents of reaction rates. 
Reaction rate uncertainty for ${\rm ^{16}O(p,\gamma)^{17}F}$ and ${\rm ^{18}O(p,\alpha)^{15}N}$ are significant to oxygen isotope ratios for all of the stellar cases studies here.
Reaction ${\rm ^{17}O(p,\alpha)^{14}N}$ competes with ${\rm ^{16}O(p,\gamma)^{17}F}$ for dominance in $^{16}$O/$^{17}$O uncertainty for the 2--3$~{\rm M}_{\sun}$ cases because of the interaction between core mixing and subsequent envelope mixing.
Of course, this result is based on the assumption that the estimated uncertaities in the reaction rates (NACRE upper and lower limits) are appropriate.

In general, reactions with large uncertainties, and reactions with slow reaction rates but large production/destruction rates and {\it any} uncertainty create the largest uncertainty in isotopic ratios.
The MC scheme demonstrated here efficiently finds the most problematic rates and provides a means of quantifying uncertainties for a particular stellar environment.

The model uncertainties calculated here are from reaction rate uncertainties only.
The total model uncertainty in oxygen istopic ratios definitly has contributions from uncertainty in mixing processes.
This problem will become more tractable as the uncertainties due to reaction rates are made smaller.

\section*{Acknowledgments}
J.\,S.\ is greatful for support from the Herzberg Institude for Astrophysics, a division of the National Reasearch Council of Canada.
F.\,H.\ appreciates  support from  \mbox{Dr.\,D.\,A.\ VandenBerg} through his
Operating Grant from the Natural Science and Engineering Research
Council of Canada.  F.\,H.\ thanks  \mbox{Dr.\ A.\,C.\ Shotter} for his
interest in this work and the hospitality at TRIUMF, Vancouver, BC.
F.\,H.\ would also like to thank \mbox{Dr.\ C.\ Iliadis} for
stimulating discussions and his hospitality at TUNL, Durham,
NC.

\onecolumn

\begin{table}
\caption{The seven different stellar cases considered with the maximum temperature achieved in the envelope during the MS. The mass coordinate of the deepest penetration of the 1dup is ${\rm M_{dup}}$ and $T_c$ is the central temperature.} \label{tbl:cases}
\begin{tabular}{ccccc}\hline
mass & metallicity & ${\rm M_{dup}}$ & max$[{\rm \log{T(M_{dup})}}]$ &max$[{\rm \log{T_c}}]$ \\ 
($M_{\sun}$) & & ($M_{\sun}$) & $(K)$& $(K)$ \\
\hline
1.0 & 0.01 & 0.23 & 7.12& -\\
1.0 & 0.02 & 0.225 & 7.11& -\\
1.5 & 0.02 & 0.25 & 7.22 & -\\
2.0 & 0.001 & 0.39 & 7.36 & 7.42\\
2.0 & 0.01 & 0.32  & 7.32 & 7.35\\
2.0 & 0.02 & 0.31  & 7.29 & 7.29\\
3.0 & 0.02 & 0.49  & 7.36 & 7.39\\
\hline
\end{tabular}
\end{table}

\begin{table}
\caption{Reactions included in the MC scheme, with nominal, lower and upper limits for reaction rates at $\log{\rm T}=7.0$ and $\log{\rm T}=7.4$.} \label{tbl:reac}
\begin{tabular}{lc}\hline
  reaction & $\log{(R_n(7.0))}~~~\log{(R_n(7.4))}$\\
           & ($sec^{-1} mol ~cm^{-3}$) \\ 
\hline

 ${\rm ^{18}O(p,\alpha)^{15}N}$   &  $  -19.98  _{  -0.70  }^{+1.13 }~~~ -11.81 _{   -0.26 }^{+    0.30}$\\ 
 ${\rm ^{15}N(p,\gamma)^{16}O}$ &   $ -20.36_{   -0.09    }^{+0.07}~~~-12.31   _{ -0.12  }^{+  0.09}$\\ 
 ${\rm ^{18}O(p,\gamma)^{19}F}$   &  $  -23.00  _{  -0.78  }^{+2.42 }~~~ -14.89 _{  -0.04  }^{+   0.17}$\\ 
 ${\rm ^{19}F(p,\alpha)^{16}O}$   & $  -23.75  _{  -0.16  }^{+0.12 }~~~ -14.42 _{  -0.16  }^{+   0.12}$\\ 
 ${\rm ^{17}O(p,\alpha)^{14}N}$   & $   -23.77 _{   -0.15  }^{+0.11 }~~~ -13.66 _{  -0.09  }^{+   0.12}$\\ 
 ${\rm ^{17}O(p,\gamma)^{18}F}$   &  $  -24.17 _{   -0.10  }^{+0.08  }~~~ -15.18  _{  -0.12  }^{+   0.12}$\\ 
 ${\rm ^{16}O(p,\gamma)^{17}F}$    &  $  -24.16 _{   -0.15   }^{+0.11}~~~ -15.41  _{  -0.15 }^{+    0.11}$\\ 
\hline
\end{tabular}
\end{table}

\begin{deluxetable}{llcc}
\tablewidth{0pt}
\tablecaption{The modeled surface oxygen isotopic ratios after 1dup, with errors from the MC simulations.\label{tbl:uncer}} 

\tablehead{
mass & metallicity, Z  &   ${\rm ^{16}O/^{17}O}$&${\rm ^{16}O/^{17}O}$\\
($M_{\sun}$) \\
}
\startdata
1.0 & 0.01 & $ 2405_{-17}^{+13}$&     $470_{-13}^{+45}$   \\
1.0 & 0.02 & $ 2410_{-16}^{+13}$&     $469_{-12}^{+46}$   \\
 1.5 & 0.02 & $1260_{-130 }^{+180}$ & $545_{-20}^{+70}$ \\
 2.0 & 0.001 &$ 177_{-25}^{+ 48}$ & $  615_{-30}^{+110}$ \\
 2.0 & 0.01 & $ 163_{-22}^{+ 53}$ & $  575_{-28}^{+100}$ \\
 2.0 & 0.02 & $ 115_{-20}^{+ 38}$ & $  565_{-20}^{+95}$ \\
 3.0 & 0.02 & $ 268_{-32}^{+ 50}$ & $  565_{-20}^{+95}$ \\
initial& &    2465 & 442 \\


\enddata
\end{deluxetable}

\begin{deluxetable}{llrrcc}
\tablewidth{0pt}
\tablecaption{Selected observations of oxygen isotope ratios. The first 8 are measured from spectroscopic observations of RGB stars. The two meteoric observations were extracted from individual dust grains that formed around AGB stars.\label{tbl:obs}} 
\tablehead{
Object & mass\tablenotemark{a} ($M_{\sun}$)& ${\rm ^{16}O/^{17}O}$ & ${\rm ^{16}O/^{18}O}$ &ref\tablenotemark{b} & [Fe/H]\tablenotemark{c}\\
}
\startdata

$\alpha$ Ari & 1   &    520$^{+100}_{-120}$ &     450  $^{   +40 }_{-    110}$  &1 &-0.221\\
$\beta$ Gem & 1.3   &   240 $^{    +50 }_{-    60 }$ &     510 $^{    +90 }_{-    110}$  &1 & -0.003\\
$\alpha$ Ser &1.4  &    300$^{     +150}_{-    150}$ &     400 $^{    +200}_{-    200}$  &1  & +0.098\\
$\beta$ UMi & 1.5  &    510 $^{    +70 }_{-    90 }$ &     440 $^{    +70 }_{-    90}$  &1 & -0.132\\
$\alpha$ Tau & 1.5  &    600 $^{    +150 }_{-   300 }$ &    475 $^{    +125 }_{-   200}$  &2 & -0.102\\
$\beta$ Peg & 1.7  &    1050 $^{   +250}_{-    500 }$ &    425 $^{    +75  }_{-   150}$  &2 &--\\
$\gamma$ Dra & 2    &    300  $^{   +75 }_{-    150 }$ &    475 $^{    +125 }_{-   200}$  &2  &-0.178\\
$\mu$ Gem & 2    &    325  $^{   +75 }_{-    150 }$ &    475 $^{    +125 }_{-   200}$  &2  &--\\
$\beta$ And & 2.5  &    155 $^{    +30 }_{-    50  }$ &    425 $^{    +75  }_{-   150}$  &2  &--\\
$\alpha$ UMa & 3    &    330  $^{   +50 }_{-    70  }$ &    600 $^{    +125 }_{-   150}$  &1 & -0.128\\
SEAL203  &--  &  355   & 498               &3   &--\\
SEAL235  &--  &  472   & 714               &3  &--\\
SEAL261  &--  &  337   & 1190              &3  &--\\
solar    &1  &  2622      &      498      &4 &0\\
\enddata
\tablenotetext{a}{ Masses have an undetermined uncertainty.}
\tablenotetext{b}{ Observations 1) at 5$\mu m$ by \citet{harris88} and 2) \citet{harris84}, 3) meteoric observations by \citet{choi:98}, 4) solar ratios by \citet{anders89}.}
\tablenotetext{c}{ From \citet{taylor}.}
\end{deluxetable}

\twocolumn
\clearpage


\begin{figure}
\resizebox{\hsize}{!}
{\includegraphics[bbllx=30pt,bblly=400pt,bburx=600pt,bbury=700pt]{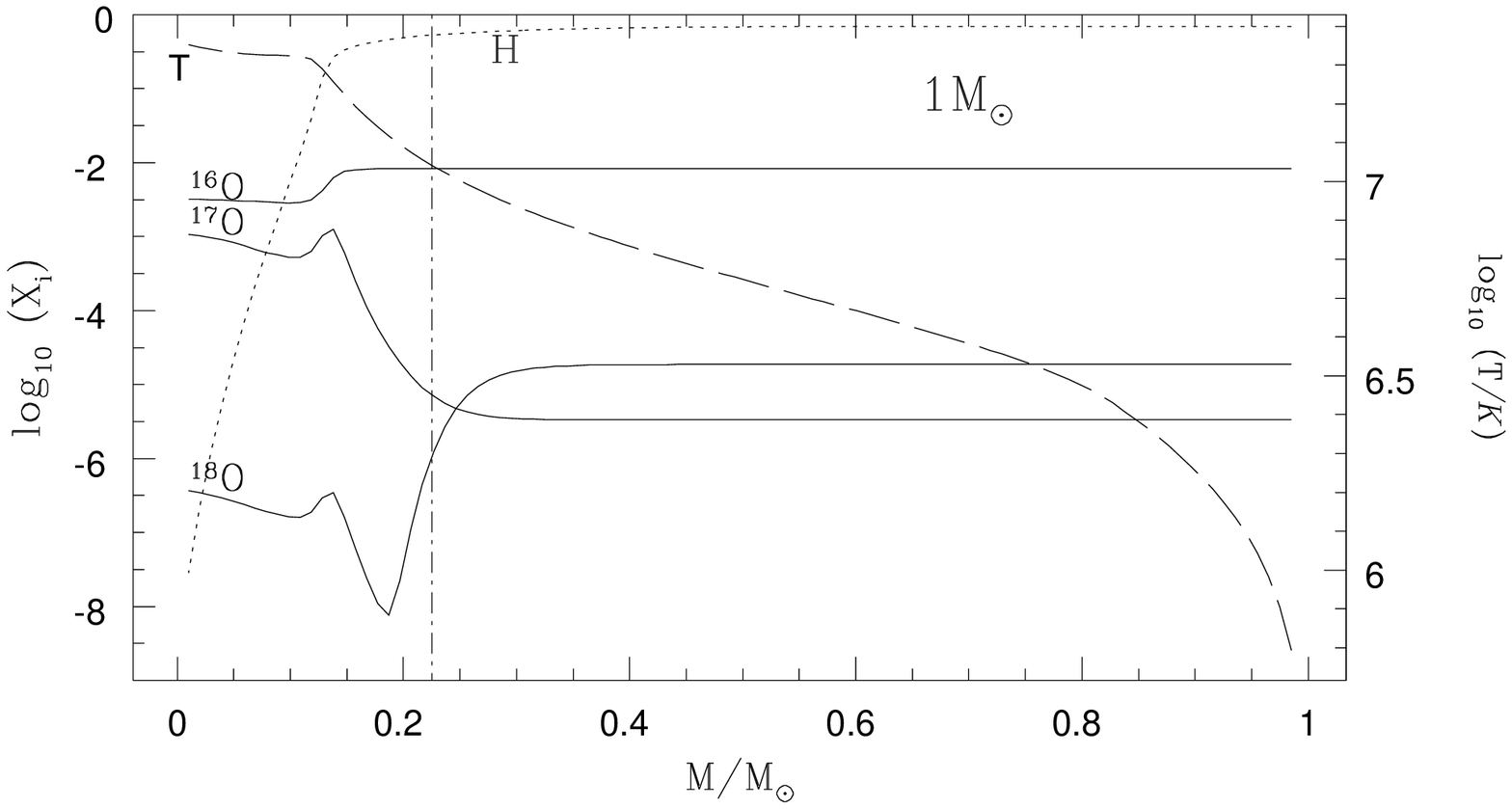}}
\end{figure}
\begin{figure}
\resizebox{\hsize}{!}
{\includegraphics[bbllx=30pt,bblly=400pt,bburx=600pt,bbury=700pt]{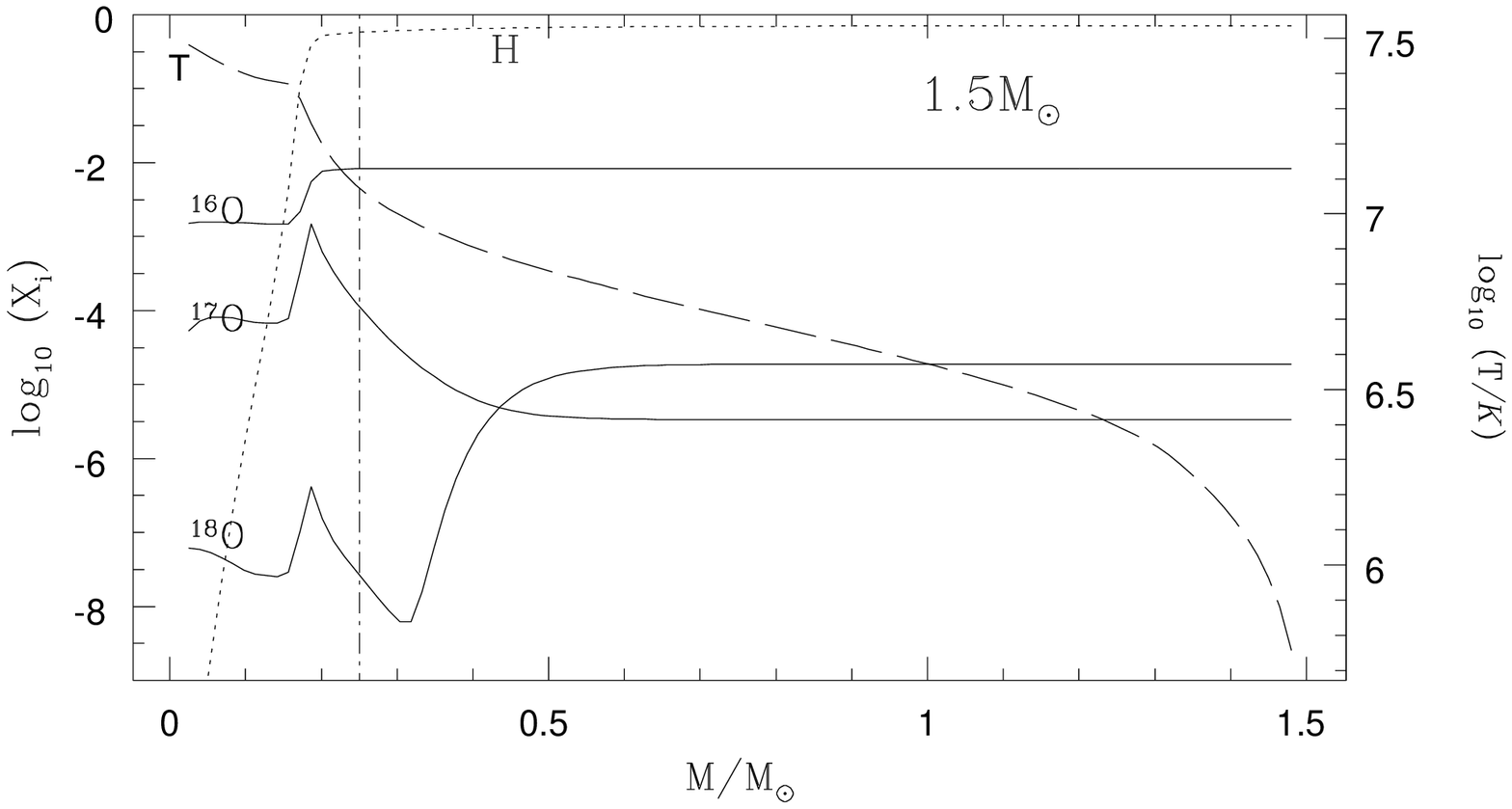}}
\end{figure}
\begin{figure}
\resizebox{\hsize}{!}
{\includegraphics[bbllx=30pt,bblly=400pt,bburx=600pt,bbury=700pt]{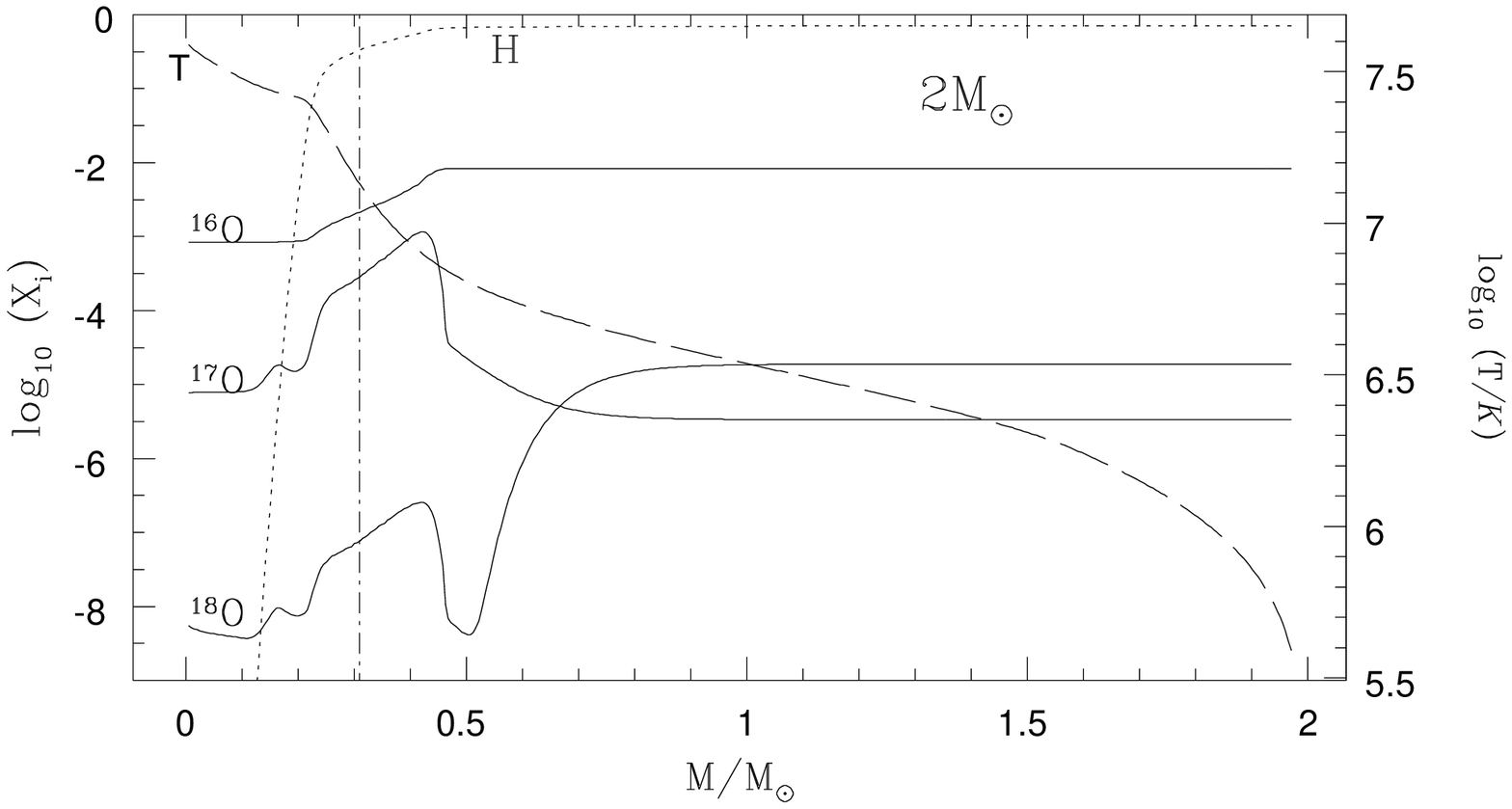}}
\end{figure}

\clearpage



\bsp

\label{lastpage}

\end{document}